\documentclass[twocolumn,           
               showpacs,            
               preprintnumbers,     
               aps,                 
               prd,                 
               a4paper,             
               groupedaddress,  
               nofootinbib,         
               tightenlines,        
               floats,floatfix      
               ]{revtex4}

\usepackage{graphicx}
\usepackage{subfigure}
\usepackage{latexsym}
\usepackage{amsmath,amssymb}        
\usepackage[draft=false]{hyperref}

\input colordvi

\begin{document}



\title{The consistency equation hierarchy in single-field inflation
models} 
\author{Marina Cort\^es}
\affiliation{Astronomy Centre, University of Sussex, Brighton BN1 9QH,
United Kingdom}
\author{Andrew R.~Liddle}
\affiliation{Astronomy Centre, University of Sussex, Brighton BN1 9QH,
United Kingdom}
\date{\today}
\pacs{98.80.Cq \hfill astro-ph/0603016}
\preprint{astro-ph/0603016}


\begin{abstract}
Inflationary consistency equations relate the scalar and tensor
perturbations.  We elucidate the infinite hierarchy of consistency
equations of single-field inflation, the first of which is the
well-known relation $A_{{\rm T}}^2/A_{{\rm S}}^2 = -n_{{\rm T}}/2$
between the amplitudes and the tensor spectral index. We write a
general expression for all consistency equations both to first and
second-order in the slow-roll expansion. We discuss the relation to
other consistency equations that have appeared in the literature, in
particular demonstrating that the approximate consistency equation
recently introduced by Chung and collaborators is equivalent to the
second consistency equation of Lidsey et al.~(1997).
\end{abstract}

\maketitle


\section{Introduction}

An important prediction of the simplest inflationary models, driven by
a single canonically-normalized scalar field, is that there should be
relations between the spectra of scalar and tensor perturbations. The
simplest such relation, usually referred to as the {\em consistency
relation}, employs the slow-roll approximation and relates the
relative amplitude of the tensor and scalar power spectra to the
tensor spectral index. Such a specific relationship, if verified
experimentally, would give powerful support to the single-field
inflationary paradigm.

In this article, we point out that this consistency relation is the
first of an infinite hierarchy of consistency relations, connecting
ever higher derivatives of the spectra. This hierarchy exists even at
lowest-order in the slow-roll approximation. That such a hierarchy
exists was first noted in the review of Lidsey et al.~\cite{LLKCBA},
but we give here for the first time explicit expressions for these
relations, both at lowest-order and next-order in slow-roll. Our
analysis is restricted to the simplest class of inflation models,
namely single-field slow-roll inflation with general relativity
assumed valid.

To some extent this exercise is an academic one, as there seems little
prospect of testing any of these relations beyond the first, and even
it is likely to prove challenging \cite{SK}. Nevertheless, these relations
offer a complete account of the connections between the two spectra,
and so any other claimed consistency relation, exact or approximate,
must follow from them if they are indeed consistency relations. In
particular we examine the relationship between our formalism and the
approximate consistency relation introduced by Chung, Shiu and Trodden
\cite{CST} and further explored by Chung and Romano \cite{CR}. We
demonstrate that it is indeed equivalent to the second consistency
equation in the hierarchy, as already given by Lidsey et
al.~\cite{LLKCBA}. 

\section{Definitions}

Following the notation of Lidsey et al.~\cite{LLKCBA}, the spectra of
scalar and tensor modes can be written
\begin{eqnarray}
\label{as}
A_{{\rm S}}(k) & \cong & \frac{4}{5 m_{{\rm Pl}}^2} \, \left[
1-(2C+1)\epsilon + C \eta\right] \, \left. \frac{H^2}{|H'|}
\right|_{k=aH} \,; \;\;\\
\label{at}
A_{{\rm T}}(k) & \cong & \frac{2}{5 \sqrt{\pi}} \, \left[
1-(C+1)\epsilon \right] \, \left. \frac{H}{m_{{\rm Pl}}}
\right|_{k=aH} \,. 
\end{eqnarray}
Here $H$ is the Hubble parameter, prime is derivative with respect to
field value $\phi$, and $C \simeq -0.73$ is a constant. The terms in
square brackets are the Stewart--Lyth slow-roll correction to the
spectrum \cite{SL}; setting the square brackets to one gives the
slow-roll result. We will use the symbol `$\cong$' to indicate
expressions as being equal within the slow-roll approximation to the
order indicated by the included terms. The first few slow-roll
parameters are defined by \cite{LPB}
\begin{eqnarray}
\epsilon &\equiv & \frac{m_{{\rm Pl}}^2}{4\pi} \, \left( \frac{H'}{H}
\right)^2 \,;\\ 
\eta &\equiv &\frac{m_{{\rm Pl}}^2}{4\pi} \, \frac{H''}{H} \,;\\ 
\xi & \equiv &\frac{m_{{\rm Pl}}^2}{4\pi} \,
\left(\frac{H'H'''}{H^2}\right)^{1/2} \,; \\
\sigma & \equiv &  \frac{m_{{\rm Pl}}^2}{4\pi} \, \left( \frac{H'^2
H''''}{H^3} \right)^{1/3}\,.
\end{eqnarray}

The wavenumber $k$ can be related to the scalar field value via the
exact relation
\begin{equation}
\frac{d\ln k}{d\phi} = \frac{4\pi}{m_{{\rm Pl}}^2} \, \frac{H}{H'} \,
(\epsilon -1) \,,
\end{equation}
where without loss of generality we have assumed $\phi$ to increase
during inflation.

The spectral indices are defined by
\begin{equation}
n_{{\rm S}}-1 \equiv \frac{d \ln A_{{\rm S}}^2}{d \ln k} \quad ; \quad
n_{{\rm T}} \equiv \frac{d \ln A_{{\rm T}}^2}{d \ln k} \,.
\end{equation}
They and their derivatives can be written in terms of the slow-roll
parameters by expressions such as
\begin{eqnarray}
n_{{\rm S}}-1 & \cong & -4\epsilon + 2\eta + \\
  & & \quad \left[-(8C+8)\epsilon^2
  + (6+10C)\epsilon\eta - 2C\xi^2 \right] \,; \nonumber \\
n_{{\rm T}} & \cong & -2\epsilon + \left[-(6+4C)\epsilon^2
  +(4+4C)\epsilon\eta 
  \right] \,; \\
\label{run}
\frac{dn_{{\rm S}}}{d\ln k} &\cong\ & -8\epsilon^2 + 10\epsilon \eta
-2\xi^2+ \\
&&\quad\left[ -(40+32C)\epsilon^3 + (60+62C)\epsilon^2\eta \right.
	\nonumber\\
&&\quad -(12+20C)\epsilon\eta^2 -(8+14C)\epsilon \xi^2 \nonumber\\
&&\left. \quad +2C \eta \xi^2 + 2C\sigma^3 \right]\, ;\nonumber\\
\frac{dn_{{\rm T}}}{d\ln k} & \cong & -4\epsilon^2+4\epsilon\eta+ \\
&&\quad\left[-(28+16C)\epsilon^3 +
  (40+28C)\epsilon^2\eta\right.\nonumber\\ 
&&\left. \quad -(8+8C)\epsilon\eta^2 -(4+4C)\epsilon\xi^2 \right]
  \,. \nonumber 
\end{eqnarray}
In each case the term enclosed in square brackets is higher order in
the slow-roll expansion, and is omitted when discussing lowest-order
results.

\section{The consistency equation hierarchy: lowest-order in slow-roll}
\label{sec: hierarchy}

In this section we restrict ourselves to the slow-roll case,
setting the square brackets in Eqs.~(\ref{as}) and (\ref{at}) equal to
one. Some simple algebra immediately leads to the standard consistency
equation
\begin{equation}
2 \frac{A_{{\rm T}}^2}{A_{{\rm S}}^2} \cong - n_{{\rm T}}
\end{equation}
Note that $n_{{\rm T}}$ is always negative by definition. This
relation was implicit in the results of Ref.~\cite{LL1}, which was the
first to write down the full slow-roll expressions, and was made
explicit and named the consistency equation in Ref.~\cite{CKLL}.

Although this is the standard form of the relation (sometimes with a
different coefficient if the spectra are defined with a different
normalization), it somewhat conceals the physical underpinning of the
consistency equation, which is that the two functions $A_{{\rm S}}(k)$
and $A_{{\rm T}}(k)$ have a common origin in the single function
$V(\phi)$, and hence must be related through elimination of $V(\phi)$
from their defining equations. This is more explicit if we write all
the scalar terms on one side and all the tensor ones on the other, to
obtain
\begin{equation}
\label{consis2}
A^{2}_{\rm S} \cong -2\frac{A^{2}_{\rm T}}{n_{\rm T}}.
\end{equation}
It is clear from this expression that specifying the tensors
completely defines the physical situation, and the corresponding
scalar spectrum can be uniquely obtained from the consistency
relation. If instead the scalars are specified, however, this is a
differential equation for the tensors whose solution yields a
one-parameter set of physical models giving that scalar spectrum and
each obeying the consistency equation.

The above equation is usually assumed to hold at one particular scale,
often combined with the somewhat inconsistent assumption that the
spectra are power-laws with different spectral indices ($n_{{\rm S}}-1
\neq n_{{\rm T}}$).  However further consistency relations can be
obtained, as first shown in Ref.~\cite{LLKCBA}, by realizing that
the consistency equation is supposed to hold on all scales. One can
proceed, for instance, by Taylor expanding both sides of
Eq.~(\ref{consis2}) in $\ln k$ about some characteristic scale
$k_0$,\footnote{In carrying out these manipulations, note that the
order-by-order slow-roll expansion is preserved both by taking
derivatives and logarithms.} giving
\begin{eqnarray}
&& A^{2}_{\rm S}+\frac{dA^{2}_{\rm S}}{d\ln k}\ln
\frac{k}{k_{0}}+\frac{1}{2} \frac{d^{2}A^{2}_{\rm S}}{d\ln
k^{2}}\ln^{2} \frac{k}{k_{0}}+\cdots \cong \quad \\ && \quad
-2\frac{A^{2}_{\rm T}}{n_{\rm T}}+\frac{d[-2\frac{A^{2}_{\rm
T}}{n_{\rm T}}]}{d\ln
k}\ln\frac{k}{k_{0}}+\frac{1}{2}\frac{d^{2}[-2\frac{A^{2}_{\rm
T}}{n_{\rm T}}]}{d\ln k^{2}}\ln^{2}\frac{k}{k_{0}}+\cdots \nonumber
\end{eqnarray}
where the expansion coefficients are all evaluated at $k_0$.
Equating the coefficients on each side gives
\begin{equation}
\label{eq:derivatives} \frac{d^{(i)}A^{2}_{\rm S}}{d\ln k^{(i)}} \cong
\frac{d^{(i)}[-2\frac{A^{2}_{\rm T}}{n_{\rm T}}]}{d\ln k^{(i)}} \,, \quad
i=0,1,\cdots \,,
\end{equation}
with both sides evaluated at some arbitrary scale $k_0$. This
represents the generic form of an infinite hierarchy of consistency
equations.

The first derivative, $i=1$, gives the lowest-order version of the
second consistency equation
\begin{eqnarray}
\label{eq:2nd}
\frac{dn_{\rm T}}{d\ln k} & \cong & 2
\frac{A^{2}_{\rm T}}{A^{2}_{\rm S}} \left[2\frac{A^{2}_{\rm
T}}{A^{2}_{\rm S}} + (n_{\rm S}-1)\right] \,. \\
\label{eq:2ndv2}
& \cong & n_{{\rm T}} \left[n_{{\rm T}}-(n_{\rm S}-1) \right] \,.
\end{eqnarray}
This equation first appeared in Ref.~\cite{KT} without being
explicitly recognized as a consistency equation, that role being
pointed out in Ref.~\cite{LLKCBA}. Eq.~(\ref{eq:derivatives}) is the
first time an explicit form for the full infinite hierarchy has been
written down.

It is possible to rewrite Eq.~(\ref{eq:derivatives}) in an
interesting alternate form using only the spectral indices
\begin{equation}
\frac{d^{(i-1)}(n_{\rm S}-1)}{d\ln k^{(i-1)}} \cong \frac{d^{(i-1)}n_{\rm
T}}{d\ln k^{(i-1)}}-\frac{d^{(i)}\ln (-n_{\rm T})}{d\ln k^{(i)}} \,, \quad
i=1,2,\cdots \,.
\end{equation}
This does not encode the normal (first) consistency relation, but does
capture all the others in quite an elegant form.

\section{The consistency equation hierarchy: next-order in slow-roll}

All of the above can readily be generalized to next-order in slow
roll by retaining the full form of Eqs.~(\ref{as}) and (\ref{at}).
The next order of the first consistency equation was first given in
Ref.~\cite{CKLL2}, and quoted in Ref.~\cite{LLKCBA} as 
\begin{equation}
\label{eq:1stn}
n_{\rm T} \cong - 2 \frac{A^{2}_{\rm
T}}{A^{2}_{\rm S}} \left[1- \frac{A^{2}_{\rm T}}{A^{2}_{\rm S}} -
(n_{\rm S}-1)\right]\,.
\end{equation}
In order to separate the scalar quantities in this expression from the
tensor ones, we write it as
\begin{equation}
\label{eq:separated}
-\frac{A^{2}_{\rm T}}{A^{2}_{\rm S}}
\frac{2}{n_{\rm T}} \cong  1-\frac{1}{2}n_{\rm T}+(n_{\rm S}-1)
\end{equation}
and use small-parameter manipulations to obtain
\begin{equation}
\label{eq:2ndsep}
A_{{\rm S}}^2 \left[1+(n_{\rm S}-1) \right] \cong -\frac{2A_{{\rm
T}}^2}{n_{\rm T}} \left[1+\frac{1}{2}n_{\rm T}\right]\,,
\end{equation}
where the scalars all stand to the left and the tensors to the
right. Note that the tensors no longer uniquely specify the scalars,
though the requirement of a subdominant next-order term (for the
expansion to make sense) will give a practically-unique scalar
spectrum for a given tensor one.

The hierarchy of consistency equations to next-order, with scalars and
tensors separated, is obtained by differentiating
Eq.~(\ref{eq:2ndsep}) repeatedly with respect to $\ln k$. For
instance, we can take Eq.~(\ref{eq:2ndv2}) to next order by
differentiating Eq.~(\ref{eq:1stn}) once to get
\begin{eqnarray}
\label{eq:2ndn}
\frac{dn_{\rm T}}{d\ln k} & \cong&
n_{\rm T} \left[n_{\rm T}-(n_{\rm S}-1)\right] \\
&& +n_{{\rm T}} \left [\frac{n_{{\rm T}}}{2} (n_{\rm T}-(n_{\rm
S}-1)) - \frac{dn_{\rm S}}{d\ln k} \right]\nonumber \,.
\end{eqnarray}
The first term on the right-hand side is of course the first-order
version of the second consistency equation.

\section{Relation to approximate consistency equations}

Since Eq.~(\ref{eq:derivatives}) and its higher-order equivalents give
a complete account of relations between the scalar and tensor spectra,
any other consistency relations claimed in the literature, approximate
or otherwise, must follow from them. One such is a relation proposed
by Chung, Shiu, and Trodden \cite{CST} and explored in detail by Chung
and Romano \cite{CR}, concerning a near coincidence of scales in
models with strong running. Another appears in Lidsey and Tavakol
\cite{LT} under the assumption of constant running. We examine each in
turn.

\subsection{Coincidence of scales}

The authors of Refs.~\cite{CST,CR} note that in models with large
running, there is an approximate coincidence of the scales where
$n_{{\rm S}}-1 = 0$ and where the tensor-to-scalar ratio reaches a
minimum. The first of those scales is denoted $k_1$, and the second
$k_2$. By definition\footnote{In parts of their paper, Chung and
Romano define scale $k_2$ as being where $\epsilon$ reaches its
extremum. Beyond the slow-roll approximation this is not quite
equivalent to our definition, which we believe is more appropriate
since $\epsilon$ is not a direct observable.}
\begin{equation}
\left. \frac{d \ln(A_{{\rm T}}^2/A_{{\rm S}}^2)}{d\ln k}
\right|_{k_2} = 0 \quad \Longrightarrow n_{{\rm S}}(k_2)-1 = n_{{\rm
T}}(k_2) \,.
\end{equation}

Since the two conditions equate $n_{{\rm S}}$ to different values, the
relation is clearly not exact.  The difference between the two scales
can be defined as $\Delta N = \ln k_2/k_1$. If we assume that the
runnings are constant, but make no assumption that the spectra arise
from inflation, there is a general expression
\begin{equation}
\Delta N = \frac{(n_{{\rm S}}-1) - n_{{\rm T}}}{dn_{{\rm T}}/d\ln k -
dn_{{\rm S}}/d\ln k} + \frac{n_{{\rm S}} - 1}{dn_{{\rm S}}/d\ln k}
\end{equation}
where the observables are evaluated at an arbitrary scale $k_0$. If we
further specialize that the expansion scale is chosen to be one of the
scales $k_1$ or $k_2$ (bearing in mind that before the fit to the data
we wouldn't know where those scales are, and that they may not lie
where the data is), this expression simplifies to
\begin{eqnarray}
\label{eq:k1}
k_1: \quad && \Delta N = -\frac{n_{{\rm T}}}{dn_{{\rm T}}/d\ln k -
dn_{{\rm S}}/d\ln k} \,;\\
\label{eq:k2}
k_2: \quad && \Delta N = \frac{n_{{\rm S}} - 1}{dn_{{\rm S}}/d\ln k}\,.
\end{eqnarray}

Two things to note about these equations are as follows. Firstly,
slow-roll inflation predicts that the two scales are far apart, not
close, since the denominator is one order higher in slow-roll than the
numerator and hence $\Delta N \sim {\cal O}(1/\epsilon)$ in the
absence of cancellations. If they are close, partial cancellations
will have allowed the running to be large while the scalar spectral
index remains close to unity (this can happen plausibly, for instance,
in running-mass inflation models \cite{runmass}). Secondly, the above
relations are {\em not} consistency relations, as no inflationary
input has been added and they are true of arbitrary spectra, not just
those tied together as inflation predicts. In particular, if $n_{{\rm
S}}$ is already measured at $k_2$, then measuring $\Delta N$ and measuring
$dn_{{\rm S}}/d\ln k$ at $k_2$ are the same thing.

The above equations can be converted into consistency equations by
substitution of the inflationary spectra, thus enforcing the relation
between tensor and scalars. For instance, doing this in
Eq.~(\ref{eq:k1}) to lowest order yields 
\begin{equation}
\label{eq:cst}
\Delta N \cong \frac{\epsilon}{\xi^2 - 4 \epsilon^2} \,,
\end{equation}
where the slow-roll parameters are evaluated at $k_1$. This is
precisely Eq.~(151) of Ref.~\cite{CST} rewritten in our
notation. Carrying out the same procedure to second-order in
Eq.~(\ref{eq:k2}) yields Eq.~(21) of Ref.~\cite{CR} (note that their
definition of the constant $C$ is different to ours).

That these relations are equivalent to the consistency equations,
specifically the second one given by Eq.~(\ref{eq:2ndv2}) or
Eq.~(\ref{eq:2ndn}), is rather subtle. Now, Eq.~(\ref{eq:cst}) is not
actually a useful form, since $\epsilon$ and $\xi$ are not directly
observable. Sufficient observables to determine them are $n_{{\rm T}}$
and $dn_{{\rm S}}/d\ln k$, bearing in mind that by definition $n_{{\rm
S}} = 1$ at the scale $k_1$ where their relation applies (and hence
$2\epsilon \cong \eta$ to the required order). This allows us to
rewrite as
\begin{equation}
\label{eq:cst2}
\Delta N \cong - \frac{n_{{\rm T}}}{n_{{\rm T}}^2 - dn_{{\rm S}}/d\ln
k} \,. 
\end{equation}
Their test therefore proposes to measure the three quantities in this
expression and verify that this relation holds.

However we know that the general expression for $\Delta N$ is
Eq.~(\ref{eq:k1}). Comparing with Eq.~(\ref{eq:cst2}), we see that
their test actually seeks to confirm that 
\begin{equation}
\frac{dn_{{\rm T}}}{d \ln k} \cong n^2_{{\rm T}} \,.
\end{equation}
This is nothing other than the second consistency equation,
Eq.~(\ref{eq:2ndv2}), evaluated at $k_1$ so that $n_{{\rm S}}-1$
vanishes. Transforming to any other scale would then give the full
version of the (lowest-order) second consistency equation.

In conclusion, while it appears that their method does not measure
$dn_{{\rm T}}/d\ln k$, in fact the measurement of $\Delta N$ along
with the other observables does so implicitly, and their test is
precisely equivalent to the second consistency equation, the
lowest-order version of which was already given in Ref.~\cite{LLKCBA}.

\subsection{Constant running}

A different relation, advertised as independent of the inflationary
potential, was given by Lidsey and Tavakol \cite{LT}. They noted that
if it were assumed that the scalar running is constant, then the
(lowest-order) equation for it, Eq.~(\ref{run}) with the square
bracket set to zero, can be written in terms of the scalar spectral
index, the tensor-to-scalar ratio, and an undetermined constant
$\tilde{c}$, eliminating the dependence on the potential. Their
Eq.~(18) reads
\begin{eqnarray}
& & \frac{A_{{\rm S}}^2}{A_{{\rm T}}^2} \exp \left[ - \frac{\left(
n_{{\rm S}}-1\right)^2}{2dn_{{\rm S}}/d\ln k} \right] \\
& & \quad - \left(
\frac{2\pi}{dn_{{\rm S}}/d\ln k} \right)^{1/2} \mathrm{erf} \left
[ \frac{n_{{\rm S}}-1}{\sqrt{2 dn_{{\rm S}}/d\ln k}} \right] =
\tilde{c} \,. \nonumber
\end{eqnarray}
As they acknowledge, in the usual interpretation where the observables
are given at a fixed (though arbitrary) expansion scale, this is not a
consistency equation as determining $\tilde{c}$ is equivalent to
determining $dn_{{\rm S}}/d\ln k$. It is further evident that it is not
a consistency equation since it does not mention the tensor spectral
index or its derivatives, whereas all members of our consistency
equation hierarchy, an exhaustive list of relations between
observables, do feature those.

They suggest that the equation can be given content by evaluating it
at two different scales, the first used to fix $\tilde{c}$ and the
second to test the relation. However this appears primarily to be a
test of the assumption of constant running, with the implications for
inflationary dynamics depending on the details of how that assumption
might fail --- typical inflation models do predict some deviation from
constant running. In any event, their relation does not follow from
the consistency equation hierarchy we have described.

\section{Conclusions}

Single-field inflation predicts not just one consistency relation, but
an infinite hierarchy, each of which can be considered at different
orders in the slow-roll expansion \cite{LLKCBA}. We have for the first
time written down explicit expressions for all these relations, and
shown how they relate to other consistency equations found in the
literature. Observed violation of these consistency relations would
exclude single-field slow-roll inflation under Einstein gravity,
pointing instead perhaps to multi-field phenomena, non-Einsteinian
gravity, or a non-inflationary origin of perturbations.

It is difficult to be optimistic about attempts to test any other than
the lowest-order version of the first consistency equation, the famous
$A_{{\rm T}}^2/A_{{\rm S}}^2 = -n_{{\rm T}}/2$ relation, which itself
is quite challenging. Song and Knox \cite{SK} have made a
comprehensive study of the ability of cosmic microwave background
experiments to test this consistency relation. They also discuss
taking that relation to next order; doing so introduces an extra
observable $n_{{\rm S}}$, which should be accurately measurable, but
current observational constraints already place us in a parameter
regime where the next-order correction should be too small to observe
due to the expected observational uncertainty on $n_{{\rm T}}$. Going
instead to the lowest-order version of the second consistency
relation, Eq.~(\ref{eq:2nd}), introduces the distinctly challenging
observable $dn_{{\rm T}}/d\ln k$. This observable is also required to
meaningfully test the coincidence of scales described in
Refs.~\cite{CST,CR}, which we have shown is equivalent to our results
and indeed those given in Ref.~\cite{LLKCBA}.


\begin{acknowledgments}
M.C.~was supported by FCT (Portugal) and A.R.L.~by PPARC (UK). We
thank James Lidsey for helpful discussions, particularly concerning
Ref.~\cite{LT}.
\end{acknowledgments}



\begin{thebibliography}{}
\bibitem{LLKCBA} J. E. Lidsey, A. R. Liddle, E. W. Kolb, E. J. Copeland,
    T. Barreiro, and M. Abney, Rev. Mod. Phys. {\bf 69}, 373 (1997),
    {\tt astro-ph/9508078}.
\bibitem{SK} Y.-S. Song and L. Knox, Phys. Rev. D{\bf 68}, 043518 (2003),
    {\tt  astro-ph/0305411}
\bibitem{CST} D. J. H. Chung, G. Shiu, and M. Trodden, Phys. Rev.
    D{\bf 68}, 063501 (2003), {\tt astro-ph/0305193}.
\bibitem{CR} D. J. H. Chung and A. E. Romano, {\tt astro-ph/0508411}.
\bibitem{SL} E. D. Stewart and D. H. Lyth, Phys. Lett. B, {\bf 302},
    171 (1993), {\tt gr-qc/9302019}.
\bibitem{LPB} A. R. Liddle, P. Parsons, and J. D. Barrow, Phys. Rev. 
    D{\bf 50}, 7222 (1994), {\tt astro-ph/9408015}.
\bibitem{LL1} A. R. Liddle and D. H. Lyth, Phys. Lett. B{\bf 291}, 391 (1992),
    {\tt astro-ph/9208007}.
\bibitem{CKLL} E. Copeland, E. W. Kolb, A. R. Liddle, and J. E. Lidsey,
    Phys. Rev. D{\bf 48}, 2529 (1993), {\tt hep-ph/9303288};
    E. J. Copeland, E. W. Kolb, A. R. Liddle, and J. E. Lidsey,
    Phys. Rev. Lett. {\bf 71}, 219 (1993), {\tt hep-ph/9304228}.
\bibitem{KT} A. Kosowsky and M. S. Turner, Phys. Rev. D{\bf 52}, 1739
    (1995), {\tt astro-ph/9504071}.
\bibitem{CKLL2}  E. Copeland, E. W. Kolb, A. R. Liddle, and J. E. Lidsey,
    Phys. Rev. D{\bf 49}, 1840 (1994), {\tt astro-ph/9308044}.
\bibitem{LT} J. E. Lidsey and R. Tavakol, Phys. Lett. B{\bf 575}, 157 
    (2003), {\tt astro-ph/0304113}.
\bibitem{runmass} E. D. Stewart, Phys. Lett. B{\bf 391}, 34 (1997),
    {\tt hep-ph/9606241}; E. D. Stewart, Phys. Rev. D{\bf 56}, 2019
    (1997), {\tt hep-ph/9703232}. 
\end{thebibliography}
\end{document}